\documentclass[twocolumn,prl]{revtex4-2}
\usepackage{amsmath}
\usepackage{amssymb}
\usepackage{graphicx}
\usepackage{dcolumn}
\usepackage{bm}
\usepackage{float}
\usepackage{mathrsfs}
\usepackage{braket}
\usepackage{hyperref}
\usepackage{mathtools}
\usepackage{comment}
\usepackage{yfonts}

\mathtoolsset{showonlyrefs=false}

\hypersetup{
 colorlinks = true,
 linkcolor = {blue},
 citecolor = {blue},
 urlcolor = {blue},
}

\usepackage[whole]{bxcjkjatype}
\usepackage[normalem]{ulem}
\usepackage[usenames,dvipsnames]{xcolor}
\usepackage{soul}
\renewcommand{\d}{\dagger}

\begin{document}
\title{Lie Algebraic Quantum Phase Reduction}
\author{Wataru Setoyama}
\email{setoyama@biom.t.u-tokyo.ac.jp}
\affiliation{
	Graduate School of Information Science and Technology,\\
	The University of Tokyo, Tokyo 113-8656, Japan
}

\author{Yoshihiko Hasegawa}
\email{hasegawa@biom.t.u-tokyo.ac.jp}
\affiliation{
 Graduate School of Information Science and Technology,\\
	The University of Tokyo, Tokyo 113-8656, Japan
}

\date{\today}

\begin{abstract}
We introduce a general framework of phase reduction theory for quantum nonlinear oscillators. 
By employing the quantum trajectory theory, we define the limit-cycle trajectory and the phase according to a stochastic Schr\"{o}dinger equation. 
Because a perturbation is represented by unitary transformation in quantum dynamics, we calculate phase response curves with respect to generators of a Lie algebra. 
Our method shows that the continuous measurement yields phase clusters and alters the phase response curves. 
The observable clusters capture the phase dynamics of individual quantum oscillators, unlike indirect indicators obtained from density operators.
Furthermore, our method can be applied to finite-level systems that lack classical counterparts.
\end{abstract}
\maketitle
\emph{Introduction.---}The last decade has witnessed a remarkable shift in the interest in synchronization, extending from classical dynamics to the quantum regime~\cite{PhysRevA.94.032336,PhysRevA.104.012410,PhysRevA.105.L020401,10.21468/SciPostPhys.12.3.097}.
Numerous studies have been reported on the synchronization of nonlinear oscillators that show quantum effects, such as spins \cite{PhysRevLett.121.063601,PhysRevLett.125.013601}, optomechanical systems \cite{PhysRevX.4.011015,Weiss_2016}, cold atoms \cite{PhysRevE.78.011108,PhysRevA.95.033808}, quantum heat engines \cite{PhysRevE.70.046110,Rezek_2006,PhysRevE.96.062120,PhysRevE.101.020201}, and (discrete or continuous) time crystals \cite{PhysRevLett.120.040404,Ke_ler_2020,doi:10.1126/science.abo3382}. 
In fact, synchronization in quantum systems is critical for considerable advances in future quantum technologies, including quantum communication and cryptography~\cite{LADD20181, doi:10.1063/5.0061478}.
For example, recent studies have shown that quantum synchronization helps addressing important security issues in quantum key distribution protocols~\cite{ Liu:19}. 
Therefore, exploring synchronization in the quantum regime holds great technological promise.
In this direction, theoretical models of limit cycles (i.e., self-sustained oscillators adaptable to weak perturbations) have been proposed in open quantum systems, such as quantum van der Pol oscillators \cite{PhysRevLett.111.234101,PhysRevLett.112.094102, walter2015quantum,PhysRevLett.123.250401,PhysRevResearch.3.013130} and spin oscillators \cite{PhysRevLett.121.053601}.
Furthermore, several experimental reports have demonstrated quantum synchronization of limit cycles in laboratory settings~\cite{PhysRevLett.125.013601,doi:10.1126/science.abo3382,PhysRevResearch.2.023026,PhysRevResearch.5.033209}.

Against this background, we propose a quantum phase-reduction theory for continuous measurement to describe quantum limit cycles in phase dynamics.
The phase reduction theory \cite{Kuramoto1984,winfree2001geometry} reduces the multidimensional dynamics of a weakly perturbed limit cycle to one-dimensional phase dynamics.
By continuously monitoring the environment to which oscillatory systems are coupled, quantum trajectories of the system come to obey a stochastic Schr\"{o}dinger equation (SSE) \cite{Gisin_1992, gardiner2004quantum, barchielli2009quantum}.
When the effect of quantum noise is sufficiently weak, these trajectories fluctuate around a deterministic trajectory.
However, since a perturbation in quantum limit-cycle dynamics differs from that in classical dynamics and is represented by a unitary transformation, we calculate the phase response to a perturbation within the Lie-algebraic framework.
Thus, we can derive a quantum phase equation from a Lindblad equation that describes a weakly perturbed dissipative system.
Note that the proposed approach reproduces the conventional phase-reduction theory in the classical limit.
Using quantum van der Pol oscillators, we show the proposal approach recovers the definitions of the limit-cycle trajectory, the phase, the perturbation, and the phase response curve (PRC) of the conventional phase-reduction theory.
Whereas Ref.~\cite{PhysRevResearch.1.033012} relies on the semiclassical approximation, we develop a fully quantum phase-reduction approach.
Thus, our approach captures the dynamics of quantum oscillators, even in the deep quantum regime and physically corresponds to the continuous measurement scheme.
Moreover, it is applicable to quantum oscillators that lack classical counterparts, such as qubits and spins.
In the quantum regime, the trajectories of quantum states are obtained by continuous measurements, where the measurement itself affect the dynamics.
Our approach captures this measurement backaction and reveals that the measurement yields phase clusters and alters the PRC in the quantum regime, through simulations of quantum van der Pol oscillators.
The resulting clusters visualize phase dynamics unique to individual quantum oscillators and cannot be captured by the indirect indicators obtained from density operators.

\emph{Derivation.---}In open quantum dynamics, quantum limit-cycle oscillators are usually described by a Lindblad equation \cite{lindblad1976generators, breuer2002theory}. Let $\rho(t)$ be a density operator at time $t$ whose time evolution is governed by
\begin{align}
    \frac{d\rho}{dt}=-i[H,\rho]+\sum_{k=1}^M \mathcal{D}[L_k]\rho,
    \label{eq:Lindblad}
\end{align}
where $H$ is a Hamiltonian operator and $L_k$ are jump operators, and $\mathcal{D}[O]$ is the dissipator defined by $\mathcal{D}[O]\rho\equiv O \rho O^{\d}-(1/2)(O^{\d}O\rho + \rho O^{\d}O)$.
To obtain a general phase-reduction approach that can be applied to quantum limit cycle models, we do not specify the jump operators $L_k$.
Note that a Lindblad equation describes a density operator, not the dynamics of the measurable quantum state.
The latter are described using quantum trajectory theory \cite{carmichael2009open}, which describes the stochastic evolution of a pure state of the system $\ket{\psi}$, obtained by continuously monitoring the environment. 
In the homodyne detection scheme, the continuous measurement can be experimentally implemented mainly by one of two approaches: detection of homodyne currents by physical detectors~\cite{gardiner2004quantum} and continuous application of weak Gaussian measurements~\cite{jacobs2006straightforward}.
In fact, quantum trajectories have been observed with various physical platforms, such as superconducting devices~\cite{murch2013observing,PhysRevX.6.011002,PhysRevLett.124.110604}, trapped ions~\cite{PhysRevLett.96.043003}, and mechanical resonators~\cite{PhysRevLett.123.163601, magrini2021real}.
In the homodyne detection scheme, the evolution can be described by the following diffusive SSE in the Stratonovich form \cite{Gisin_1992, gardiner2004quantum, barchielli2009quantum, supply}.\nocite{10.1162/neco.1996.8.5.979,PhysRevA.91.062308}
\begin{align}
    d\ket{\psi} =& \left[-iH_{\mathrm{eff}}+\sum_{k=1}^M \frac{1}{2}\braket{L_k^{\d}L_k} +\braket{X_k}\left(L_k-\frac{\braket{X_k}}{2}\right)\right.\nonumber\\
    &\left.+\frac{1}{4}\left(-2L_k^2+\braket{L_k^2}+\braket{L_k^{\d 2}}\right)\right]\ket{\psi}dt\nonumber\\
    &+\sum_{k=1}^M \left(L_k - \frac{\braket{X_k}}{2}\right)\ket{\psi}\circ dW_k(t),
    \label{eq:SSE}
\end{align}
where $\circ$ denotes the Stratonovich calculus, $H_{\mathrm{eff}}\equiv H-(i/2)\sum_{k=1}^ML_k^{\d}L_k$ is a non-Hermitian operator (i.e., an effective Hamiltonian), and $X_k \equiv L_k+L_k^{\d}$ are quadratures of the system. Here, $\braket{O}$ denotes the expectation value of $O$ with respect to state $\ket{\psi}$, i.e., $\braket{O}\equiv \bra{\psi}O\ket{\psi}$. Random variables $dW_k$ are Wiener increments that satisfy $\mathbb{E}[dW_k]=0$, and $\mathbb{E}[dW_k^2]=dt$, where $\mathbb{E}[\cdot]$ denotes the average over all possible trajectories. The homodyne currents $J_k$ are defined as $J_k(t)\equiv\braket{X_k}+\xi_k(t)$, where $\xi_k(t)\equiv dW_k/dt$. In general, a limit-cycle trajectory in quantum dynamics and the phase along it is not well defined. In the classical stochastic dynamics of limit cycles, a stochastic differential equation is represented by adding noise terms to a given deterministic differential equation \cite{GOLDOBIN2005126,PhysRevLett.98.184101,PhysRevLett.101.154101,PhysRevLett.102.194102}. In contrast, quantum dynamics are stochastic in nature and the deterministic equation is not given. To realize a quantum phase reduction, the deterministic limit cycle and the phase along it should be defined. The classical deterministic limit-cycle dynamics corresponds to an equation obtained by removing noise terms from a stochastic differential equation in the Stratonovich form. As an analog of classical cases, we propose here to remove noise terms from an SSE in the Stratonovich form and define the resulting equation as the deterministic limit-cycle dynamics:
\begin{align}
d\ket{\psi} =& \left[-iH_{\mathrm{eff}}+\sum_{k=1}^M \frac{1}{2}\braket{L_k^{\d}L_k} +\braket{X_k}\left(L_k-\frac{\braket{X_k}}{2}\right)\right.\nonumber\\
    &\left.+\frac{1}{4}\left(-2L_k^2+\braket{L_k^2}+\braket{L_k^{\d 2}}\right)\right]\ket{\psi}dt.
    \label{eq:limcyc}
\end{align}
An SSE is usually represented and calculated in the Ito form for computational and statistical convenience. It is worth emphasizing that noise terms should be removed from an SSE in the Stratonovich interpretation, rather than in the Ito interpretation for the following reasons. The first is related to the chain rule of differentiation calculation. In fact, the phase reduction requires a coordinate transformation between a state vector and a phase coordinate. The transformation is performed via the chain rule of differentiation, which holds only in the Stratonovich form (not in the Ito form). 
The second reason is related to norm preservation. 
To ensure that the limit cycles represent physically observable trajectories of pure states, it is essential to satisfy norm preservation.
Note that the norm of Eq.~\eqref{eq:limcyc} is preserved, $d\|\psi\|=0$, where $\|\psi\|\equiv\sqrt{\braket{\psi|\psi}}$.
Therefore, Eq.~\eqref{eq:limcyc} stands on its own as pure-state dynamics, which is not the case for the Ito interpretation. 
Even when considering an arbitrary stochastic calculus, the norm preservation is satisfied only in the Stratonovich calculus~\cite{supply}.
It is a nontrivial property that an SSE with noise terms removed also stands as pure-state dynamics because, unlike the case for classical dynamics, the deterministic dynamics Eq.~\eqref{eq:limcyc} is not given. 

When Eq.~\eqref{eq:limcyc} satisfies $\lim_{t\to\infty}|\braket{\psi(t)|\psi(t+T)}|=1$ for a period $T$, $\ket{\psi}$ has a limit-cycle solution $\ket{\psi_0}$ to which $\ket{\psi}$ converges. Since $U(1)$ has no physical effect on the state $\ket{\psi}$ \cite{weyl1950theory}, the $U(1)$ transformation has no effect on the phase.
We define the phase on a quantum limit cycle using the deterministic trajectory $\ket{\psi_0}$. There are several schemes for the phase reduction in classical stochastic systems \cite{PhysRevLett.98.184101,PhysRevLett.101.154101,https://doi.org/10.48550/arxiv.0812.3205}. For simplicity, we derive the phase equation by following the procedure in \cite{PhysRevLett.98.184101}. The phase $\theta$ is defined along the limit-cycle solution $\ket{\psi_0}$ using Eq.~\eqref{eq:limcyc} as to change at a constant frequency $\omega=2\pi/T$. Furthermore, by virtue of the convergence to the limit-cycle solution $\ket{\psi_0}$, the phase $\theta$ outside of it is defined by an isochron under Eq.~\eqref{eq:limcyc} as $\Theta\left(\ket{\psi(t)}\right)\equiv \Theta\left(\lim_{n\to\infty}\ket{\psi(t+nT)}\right)$, where the phase function $\Theta\left(\ket{\psi}\right)$ represents the phase at the state $\ket{\psi}$. Here, we assume that the perturbation is sufficiently weak, i.e., the state $\ket{\psi}$ is in the vicinity of the limit-cycle solution $\ket{\psi_0}$.

It should be mentioned that our definition of the PRC differs from that of the classical counterpart. While the state is defined in the Euclidean space for the classical limit cycle, the unitary group in the Hilbert space defines the state of a quantum limit cycle. Therefore, the corresponding bases are the generators of the unitary group $U(N)$ \cite{fecko2006differential}. They can be decomposed into the generators of $U(1)$ and those of the special unitary group $SU(N)$. $U(1)$ represents the scalar multiplication, while $SU(N)$ is a unitary group with a determinant $\det[U]=1$. For example, the generators of $SU(2)$ correspond to Pauli matrices. By the definition of the phase, $U(1)$ has no effect on it. Thus, only $SU(N)$ should be considered for the phase dynamics. In quantum limit cycles, the perturbation is represented by an infinitesimal unitary transformation and the PRC is calculated for it. Based on a Lie algebra, an arbitrary infinitesimal unitary transformation is represented by the Taylor expansion as $U\ket{\psi}=\exp\left(\sum_{l=1}^{N^2-1} -ig_l E_l - ig_0 I\right)\ket{\psi} \simeq \ket{\psi}-\sum_{l=1}^{N^2-1} ig_l E_l\ket{\psi}-ig_0\ket{\psi}$, where $E_l$ are generators of $SU(N)$, $I$ is the identity matrix, and real coefficients $g_l$ satisfy $|g_l| \ll 1$. The PRCs for the generators $E_l$ are represented as
\begin{align}
    Z_l(\theta)\equiv \lim_{g_l\rightarrow 0}\frac{\Theta(\exp(-ig_l E_l)\ket{\psi_0(\theta)})-\Theta(\ket{\psi_0(\theta)})}{g_l},
    \label{eq:phaseresponse}
\end{align}
where $\ket{\psi_0(\theta)}$ represents the state $\ket{\psi}$ on the limit-cycle solution $\ket{\psi_0}$ with phase $\theta$. Equation~\eqref{eq:phaseresponse} describes the partial derivative with respect to a unitary transformation by generator $E_l$. This formulation defines the quantum PRC. 
For the case of high-dimensional systems, e.g., semiclassical systems, PRCs with respect to $N^2-1$ generators of a Lie algebra demand large computational resource. In such a case, we can calculate PRC either by a direct method with respect to an arbitrary Hamiltonian or an adjoint method in the Euclidean space~\cite{supply}.
While the SSE [Eq.~\eqref{eq:SSE}] and Eq.~\eqref{eq:limcyc} are described as non-Hermitian dynamics, owing to their non-linearity, they can also be represented as Hermitian dynamics~\cite{supply}.
Thus, the stochastic terms in an SSE can be represented by traceless Hermitian operators as $d\ket{\psi}=-i\sum_{k=1}^{M}H_k\ket{\psi}\circ dW_k$~\cite{supply}, where traceless Hermitian operators $H_k$ are defined by
\begin{align}
    H_k \equiv i(L_k-\braket{L_k})\ket{\psi}\bra{\psi}+\mathrm{H.c.}
    \label{eq:Hk}
\end{align}
Due to the trace-orthogonal property of the Lie algebra, traceless Hermitian operators $H_k$ can be decomposed into a linear combination of $SU(N)$ generators as $H_k=\sum_{l=1}^{N^2-1}g_{k,l}E_l$, where the coefficients $g_{k,l}$ are defined by $g_{k,l}\equiv \mathrm{Tr}[H_{k} E_l]$.
Therefore, the following quantum phase equation is derived from the chain rule
\begin{align}
    \frac{d\theta}{dt}=\omega+\sum_{k=1}^M\sum_{l=1}^{N^2-1} Z_l(\theta)g_{k,l}(\theta)\circ \xi_k(t),
    \label{eq:PE_str}
\end{align}
where $g_{k,l}(\theta)$ is evaluated at $\ket{\psi}=\ket{\psi_0(\theta)}$ on the limit cycle. The phase equation \eqref{eq:PE_str} in the Stratonovich form can be converted into an equivalent equation in the Ito form \cite{gardiner2004handbook}
\begin{align}
     \frac{d\theta}{dt}=\omega+\frac{1}{2}\sum_{k=1}^M \frac{dY_k(\theta)}{d\theta}Y_k(\theta) + \sum_{k=1}^M Y_k(\theta) \xi_k(t),
     \label{eq:PE_ito}
\end{align}
where $Y_k(\theta)=\sum_{l=1}^{N^2-1} Z_l(\theta)g_{k,l}(\theta)$. 
As long as the quantum dynamics is represented by the SSE [Eq.~\eqref{eq:SSE}], arbitrary weak perturbations can be considered in our framework~\cite{supply}. 
In the following, we shall elaborate on the difference between our approach and that in Ref.~\cite{PhysRevResearch.1.033012}, which is the extant phase-reduction approach for quantum systems.
In a semiclassical approximation, Ref.~\cite{PhysRevResearch.1.033012} reduces quantum dynamics to classical one based on a quasi-probability distribution~\cite{carmichael1999statistical,gardiner2004quantum}, and applies the conventional phase-reduction theory to it. In contrast, based on a Lie algebra, our approach proposes the original framework of phase reduction theory directly applicable to the pure state $\ket{\psi}$ of quantum limit cycles. 
To explain the difference in detail, we examine the quantum van der Pol oscillator defined by
\begin{align}
    \dfrac{d\rho}{dt}=-i[H,\rho]+\gamma_{1g}\mathcal{D}[a^{\d}]\rho+\gamma_{1d}\mathcal{D}[a]\rho+\gamma_{2d}\mathcal{D}[a^2]\rho,
\label{eq:qvdP}
\end{align}
where $H=a^{\d}a$ is the Hamiltonian and $a$ and $a^{\d}$ are annihilation and creation operators, respectively. 
The quantum van der Pol model describes the limit-cycle dynamics at a quantum scale. In quantum systems, the measurement outcomes are stochastic in nature. Thus, the position $x=(1/\sqrt{2})(a+a^{\d})$ and the momentum $p=-(i/\sqrt{2})(a-a^{\d})$ are evaluated through their expectation values as $\braket{x}_{\rho}$ and $\braket{p}_{\rho}$, respectively, where $\braket{O}_{\rho}\equiv \mathrm{Tr}[O\rho]$. 
In the classical limit, $\braket{a^{\d}a}_{\rho}\gg 1$ (i.e., the system is at a macroscopic scale), 
Eq.~\eqref{eq:qvdP} gives the equation as follows:
\begin{align}
    \frac{d\alpha}{dt} = -i\alpha+\frac{\epsilon}{2}\alpha -\gamma_{2d}|\alpha|^2\alpha,
    \label{eq:alpha}
\end{align}
where $\alpha \equiv (\braket{x}_{\rho}+i\braket{p}_{\rho})/\sqrt{2}$
and $\epsilon\equiv \gamma_{1g}-\gamma_{1d}$ corresponds to the difference between one-particle gain and loss rates.
Differentiating real part of Eq.~\eqref{eq:alpha} with respect to time and substituting imaginary part of Eq.~\eqref{eq:alpha} into it,
the classical van der Pol model is recovered up to $O(\epsilon^2)$ as $\braket{\Ddot{x}}_{\rho}+\braket{x}_{\rho}=\epsilon\left\{1-(\braket{x}_{\rho}^2+\braket{\Dot{x}}_{\rho}^2)/{A_c}^2\right\}\braket{\Dot{x}}_{\rho}+O(\epsilon^2)$, where $A_c\equiv \sqrt{\epsilon/\gamma_{2d}}$~\cite{PhysRevResearch.3.013130}.
For the semiclassical approximation, the previous work in \cite{PhysRevResearch.1.033012} can be applied only to systems near the classical limit $\gamma_{1g}\gg \gamma_{2d}$. In contrast, our approach can be applied to an arbitrary regime, including the deep quantum regime $\gamma_{2d}\gg \gamma_{1g}$. 
Similarly, our approach differs from Ref.~\cite{kato2021enhancement}, which is a feedback control scheme to enhance synchronization by applying the semiclassical phase reduction to a homodyne detection scheme.

Thus far, we have been concerned with regimes ranging from the semiclassical to the quantum regime. Historically, the phase reduction theory was demonstrated in the context of classical deterministic dynamics. In the following, we show that our approach reduces to the conventional phase-reduction theory in the classical limit. 
In the classical limit, the state $\ket{\psi}$ is considered coherent and satisfies $a\ket{\psi}=\alpha\ket{\psi}$, $a^{\d}\ket{\psi}=\alpha^{*}\ket{\psi}+\ket{x}$, and $|\alpha| \gg 1$, where $\ket{x}\equiv a^{\d}\ket{\psi}-\alpha^{*}\ket{\psi}$. 
Substituting these conditions into Eq.~\eqref{eq:SSE}, we obtain 
\begin{align}
    \frac{d\alpha}{dt} = \left[-i\alpha+\frac{\epsilon}{2}\alpha -\gamma_{2d}|\alpha|^2\alpha\right] + \sqrt{\gamma_{1g}}\circ \xi(t).
    \label{eq:classicallimit}
\end{align}
In Eq.~\eqref{eq:classicallimit}, the deterministic term, which equals Eq.~\eqref{eq:alpha}, is $O(|\alpha|^2\alpha)$ whereas the stochastic term is $O(1)$. 
Hence, in the classical limit, the dynamics can be considered as deterministic and its limit cycle is equivalent to the classical one.
In the classical limit, the proposed and semiclassical methods give the same limit cycle not only for quantum van der Pol but also in general cases~\cite{supply}.
Moreover, this equivalence applies also to the perturbation and phase response. 
In the conventional method, the perturbation is represented by basis vector $dx$ in the Euclidean space. It can be reproduced by the momentum operator $p$ in the Hilbert space as follows: By the unitary perturbation $d\ket{\psi}/dt=-ip\ket{\psi}$, the derivative of the expectation value of the position is unity, i.e., $d\braket{x}/dt=-i\braket{[x,p]}=1$. The same argument holds for $dp$.
Because the same perturbation can be reproduced, the PRC in the conventional method can also be replicated similarly in the classical limit. The conventional PRC is defined as $Z_{\mathrm{cl}}(\theta)\equiv  [\Theta(\alpha(\theta) + dx)-\Theta(\alpha(\theta))]/dx$, and it can be reconstructed by the unitary transformation as $Z(\theta)=\lim_{h\to 0}[\Theta(\exp(-ihp)\ket{\psi_0(\theta)})-\Theta(\ket{\psi_0(\theta)})]/h$.
\begin{figure}
\includegraphics[width=8.5cm]{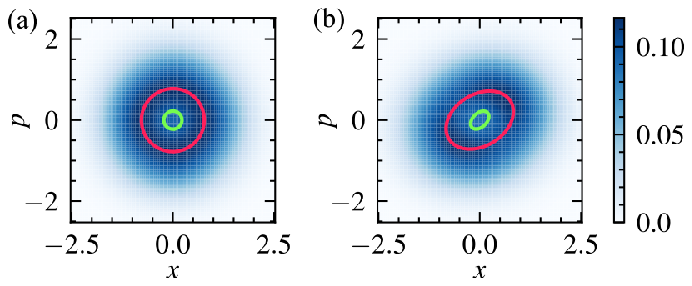}
\caption{Wigner function and limit-cycle trajectories of quantum and semiclassical phase-reduction approaches in the two-parameter settings quantum regime, (a) $(\Delta$, $\Omega$, $\eta\mathrm{e}^{i\lambda}$, $\gamma_{1g}$, $\gamma_{1d})/\gamma_{2d}=(1$, $0$, $0$, $0.1$, $0)$ and (b) $(\Delta$, $\Omega$, $\eta\mathrm{e}^{i\lambda}$, $\gamma_{1g}$, $\gamma_{1d})/\gamma_{2d}=(1$, $0$, $-0.2$, $0.1$, $0)$. Color intensity is proportional to the quasi-probability of the Wigner function. The limit-cycle trajectory of the quantum phase reduction (red line) passes through the high quasi-probability region of the Wigner function, while that of the semiclassical phase reduction (green line) does not. The fidelity $F(\rho_1,\rho_2)\equiv\mathrm{Tr}\left[\sqrt{\sqrt{\rho_1}\rho_2\sqrt{\rho_1}}\right]^2$ between the true density operator and those reconstructed from the phase distribution are (a) $0.958$, (b) $0.979$ in our method, and (a) $0.812$ in the semiclassical method.}\label{fig:Wigner}      
\end{figure}
\begin{figure}
\includegraphics[width=8.5cm]{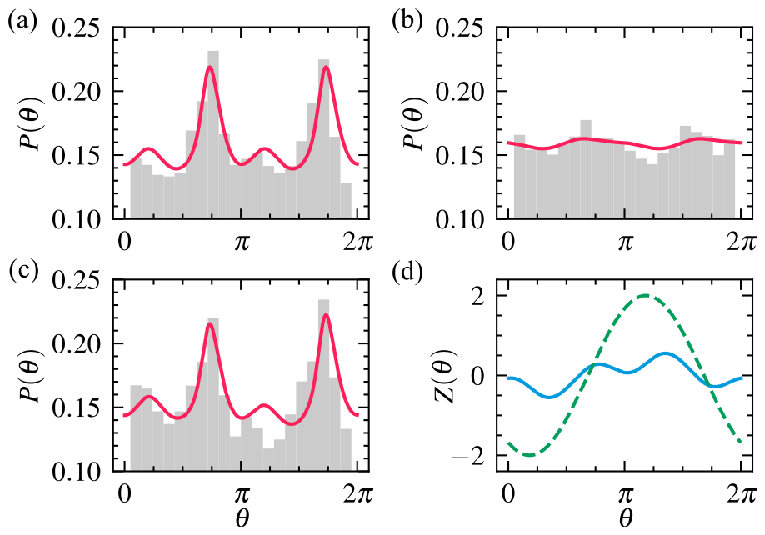}
\caption{Phase distribution $P(\theta)$ in steady state and PRC $Z(\theta)$. (a), (b), and (c) Phase distribution (a) in quantum regime, (b) in deep quantum regime, and (c) subjected to harmonic drive as weak perturbation. (d) PRCs with respect to harmonic drive. The gray histograms are computed from SSE simulations and the red lines are computed from simulations of the proposed phase equation for (a), (b), and (c). 
The solid blue line is obtained from the proposed method and the dashed green line is obtained from the semiclassical method for (d).
The strength of the weak perturbation is $\Omega_p=0.05$ for (c). The parameters are  $(\Delta$, $\Omega$, $\eta\mathrm{e}^{i\lambda}$, $\gamma_{1g}$, $\gamma_{1d})/\gamma_{2d}=(1$, $0$, $0$, $0.5$, $0)$ for (a), (c), and (d), $(\Delta$, $\Omega$, $\eta\mathrm{e}^{i\lambda}$, $\gamma_{1g}$, $\gamma_{1d})/\gamma_{2d}=(1$, $0$, $0$, $0.1$, $0)$ for (b).}\label{fig:dist_prc}
\end{figure}
\begin{figure*}[t]
    \centering
    \includegraphics[width=18cm]{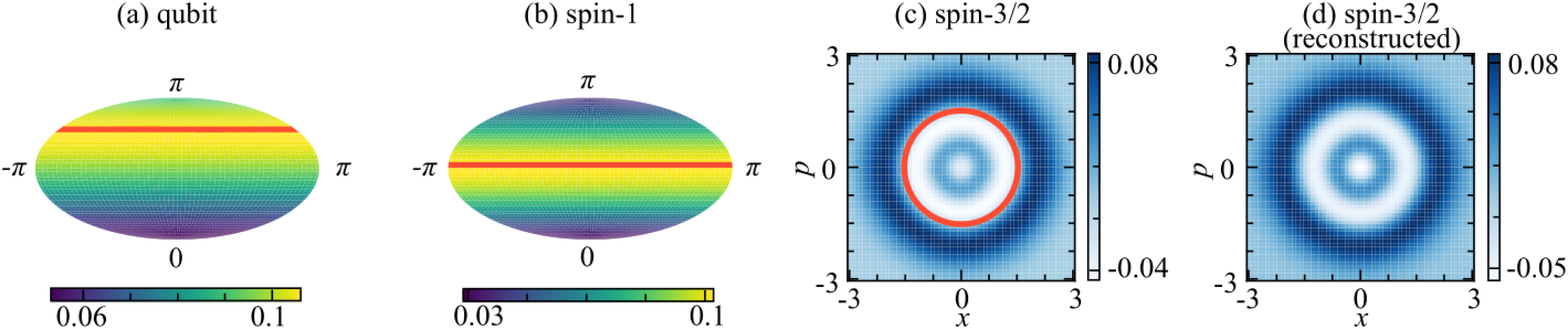}
    \caption{Expectation values on limit-cycle trajectories and quasiprobability distributions of (a) two-level systems~\cite{PhysRevA.101.062104}, (b) spin-1 oscillators~\cite{PhysRevLett.121.053601}, and (c) spin-3/2 oscillators; (d) reconstructed quasiprobability distribution for spin-3/2 spins from phase distribution. The quasiprobability distribution is Husimi-Q function for (a) and (b) and Wigner function for (c) and (d). The set of observables evaluated on limit-cycle trajectories are spin operators for (a) and (b) and position and momentum operators for (c).}
    \label{fig:finite}
\end{figure*}

\emph{Example.---}As an example, we consider the quantum van der Pol oscillators [Eq.~\eqref{eq:qvdP}] in a rotating frame~\cite{PhysRevLett.112.094102}, where $H=-\Delta a^{\dagger}a + i\Omega(a^{\dagger}-a) +i\eta(a^2\exp(-i \lambda)-a^{\dagger 2}\exp(i \lambda))$ is the Hamiltonian, $\Delta=\omega_d-\omega_0$ is the detuning between the system's natural frequency $\omega_0$ and a harmonic drive frequency $\omega_d$, $\Omega$ is the strength of the harmonic drive, and $\eta$ and $\lambda$ are the strength and phase of squeezing, respectively.
In a rotating frame, the system rotates with a harmonic drive frequency $\omega_d$. 
First, we numerically validate the accuracy of the approximation by comparing the derived phase equation to the semiclassical phase equation.  
Figure~\ref{fig:Wigner} shows the Wigner function in the steady state and the limit-cycle trajectory of each phase-reduction method in the quantum regime. 
We calculate the reconstructed density operator $\rho_{\mathrm{re}}\equiv\int d\theta P(\theta)\ket{\psi_0(\theta)}\bra{\psi_0(\theta)}$, where $P(\theta)$ is a probability density function of the phase $\theta$, for each phase-reduction method. Furthermore, we compare their fidelity level to those of the true density operator~\cite{PhysRevResearch.1.033012}. 
Our method provides a better approximation than the semiclassical method (see the caption of Fig.~\ref{fig:Wigner} for details), because it reduces a pure state to the phase without the semiclassical approximation. Note that we cannot calculate fidelity for the semiclassical method in Fig.~\ref{fig:Wigner}(b) because diffusion matrices of a semiclassical Langevin equation are not positive-semidefinite in some points.

Next, we investigate the effect of the measurement and the harmonic drive on quantum synchronization in the quantum regime.
In contrast to classical dynamics, the measurement affects the system's trajectory in the quantum regime. 
For brevity of expression, we here approximate the quantum van der Pol oscillator by limiting the bosonic Fock state to the lowest $N$ levels~\cite{PhysRevLett.123.250401}. In Fig.~\ref{fig:dist_prc}, $N=6$ for (a), (c), and (d), and $N=4$ for (b).
In the quantum regime, the proposed method yields a good approximation, as shown in Fig.~\ref{fig:dist_prc}(a).
Although clusters diminish for the deep quantum regime in Fig.~\ref{fig:dist_prc}(b), in both cases, the measurement generates the clusters in the rotating frame in Figs.~\ref{fig:dist_prc}(a) and (b).
Furthermore, as a weak perturbation Hamiltonian $H_p=i\Omega_p(a^{\dagger}-a)$, a harmonic drive is added to the Hamiltonian $H$ and enhances synchronization in Fig.~\ref{fig:dist_prc}(c), where $\Omega_p$ is the strength of the weak perturbation. 
In Fig.~\ref{fig:dist_prc}(d), the PRC of the proposed method appears distorted due to the measurement, unlike that of the semiclassical method, which exhibits a sinusoidal wave pattern. Moreover, in the quantum regime, the limit cycle shrinks due to the classical approximation, resulting in larger amplitude for the PRC in the semiclassical method.
Some indicators have been proposed as signatures of quantum synchronization, such as mutual information \cite{PhysRevA.91.012301}, quantum discord \cite{PhysRevA.85.052101,Zhu_2015}, and entanglement \cite{PhysRevE.89.022913}. 
The observable clusters describe the synchronization dynamics of individual oscillators under the measurement backaction, unlike the indirect indicators obtained from a density operator.

Additionally, we demonstrate the applicability of our method to finite-level systems. 
Qubits and spins hold a central place in the field of quantum synchronization; however, they lack classical limit-cycle counterparts. 
We apply the proposed method to two-level systems~\cite{PhysRevA.101.062104}, spin-1 oscillators~\cite{PhysRevLett.121.053601} at finite temperature, and spin-3/2 oscillators, and derive the phase equations~\cite{supply}.
Figure~\ref{fig:finite} displays the quasiprobability distributions for each model and the expected values of observables evaluated on the limit cycles.
As shown in Figs.~\ref{fig:finite}(a) and (b), the trajectories for the qubit and spin-1 pass through regions of high probability. 
Since we plot expectation values, as shown in Fig.~3(c), the trajectories for spin-3/2 pass between two regions of high probability. 
Yet, the Wigner distribution is reconstructed from the phase distribution with fidelity $F=0.998$ in Fig.~\ref{fig:finite}(d).

\emph{Conclusion.---}In this Letter, we proposed a quantum phase-reduction formulation of a Lindblad equation in a continuous measurement scheme.
We consider the case of synchronization among multiple quantum oscillators with the proposed method in~\cite{setoyama2023lie}.
\begin{acknowledgments}
This work was supported by JSPS KAKENHI Grant No. JP22H03659.
\end{acknowledgments}

%

\end{document}


\title{Supplementary Material for \\``Lie Algebraic Quantum Phase Reduction''}
\author{Wataru Setoyama}
\email{setoyama@biom.t.u-tokyo.ac.jp}
\affiliation{Department of Information and Communication Engineering, Graduate
School of Information Science and Technology, The University of Tokyo,
Tokyo 113-8656, Japan}
\author{Yoshihiko Hasegawa}
\email{hasegawa@biom.t.u-tokyo.ac.jp}
\affiliation{Department of Information and Communication Engineering, Graduate
School of Information Science and Technology, The University of Tokyo,
Tokyo 113-8656, Japan}

\maketitle
This supplementary material describes the calculations introduced in the main text. Equation and figure numbers are prefixed with S (e.g., Eq.~(S1) or Fig.~S1). Numbers without this prefix (e.g., Eq.~(1) or Fig.~1) refer to items in the main text.

\tableofcontents

\section{Stochastic calculus}
We introduce the conversion of stochastic differential equations between the two typical stochastic calculi, the Ito and Stratonovich calculi. Furthermore, we apply this conversion to the stochastic Schr\"{o}dinger equation (SSE).

\subsection{Definitions of stochastic calculi}\label{appendix:stochastic calculus}
Typically, a stochastic differential equation is represented by two different forms, the Ito and Stratonovich forms, which are given as follows:
\begin{align}
    f(t)dW(t)&\equiv f(t)\left(W(t+dt)-W(t)\right),\\
     f(t)\circ dW(t)&\equiv f\left(t+\frac{dt}{2}\right)\left(W(t+dt)-W(t)\right),
\end{align}
where $W(t)$ is a Wiener process. In the Stratonovich interpretation, a calculus is performed at the midpoint of the interval $[t, t+dt]$. Moreover, the two forms can be converted to each other using the following transformation:
\begin{align}
    f(t)\circ dW(t)=f(t)dW(t)+\frac{1}{2}df(t)dW(t).
\label{eq:conversion}
\end{align}
The conversion of Eq.~\eqref{eq:conversion} is performed according to the Ito rule:
\begin{align}
    dW(t)^2&=dt,\\
    dW(t)dt&=0.
\end{align}

\subsection{Ito and Stratonovich conversion of SSE}\label{appendix:conversion}
A diffusive SSE is represented by the Ito form \cite{Gisin_1992, gardiner2004quantum}:
\begin{align}
    d\ket{\psi} =& dt\left[ -iH_{\mathrm{eff}}+\sum_{k=1}^M \frac{\braket{X_k}}{2}\left(L_k - \frac{\braket{X_k}}{4}\right)\right]\ket{\psi}\nonumber\\
    &+\sum_{k=1}^M dW_k(t)\left(L_k - \frac{\braket{X_k}}{2}\right)\ket{\psi}.
    \label{eq:SSE_Ito}
\end{align}
where $H_{\mathrm{eff}}\equiv H-(i/2)\sum_{k=1}^M L_k^{\d}L_k$ is an effective Hamiltonian and $X_k\equiv L_k+L_k^{\d}$ are quadrature operators of the system. According to Section \ref{appendix:stochastic calculus}, Eq.~\eqref{eq:SSE_Ito} can be converted to the equivalent Stratonovich form as follows:
\begin{align}
    d\ket{\psi} =& \left[ -iH_{\mathrm{eff}}+\sum_{k=1}^M \frac{1}{2}\braket{L_k^{\d}L_k} +\braket{X_k}\left(L_k-\frac{\braket{X_k}}{2}\right)+\frac{1}{4}\left(-2L_k^2+\braket{L_k^2}+\braket{L_k^{\d 2}}\right)\right]\ket{\psi}dt\nonumber\\
    &+\sum_{k=1}^M \left(L_k - \frac{\braket{X_k}}{2}\right)\ket{\psi}\circ dW_k(t).
    \label{eq:SSE_Str}    
\end{align}
The conversion from Eq.~\eqref{eq:SSE_Ito} to Eq.~\eqref{eq:SSE_Str} is performed according to the multivariate Ito rule \cite{gardiner2004handbook}:
\begin{align}
        dW_m(t)dW_n(t)&=\delta_{m,n}dt,\\
    dW_m(t)dt&=0,
\end{align}
where $\delta_{m,n}$ is the Kronecker's delta.

\section{Generators of Lie algebra}
In the main text, we define infinitesimal unitary transformation as a perturbation of quantum limit cycles in Hilbert space.
Arbitrary unitary transformation $U$ can be represented by an Hermitian operator $H$ as $U\equiv \exp(-iH)$, and $N$ dimensional Hermitian operator $H$ can be decomposed into generators of special unitary group $SU(N)$. Here, we introduce generators of $SU(3)$ and $SU(N)$.  

\subsection{Generators of $SU(3)$}\label{appendix:Lie}
The generators of $SU(3)$ correspond to Gell-Mann matrices $\{\lambda_l\}$. 
\begin{align}
\lambda_1&=    \begin{bmatrix}
0 & 1 & 0 \\
1 & 0 & 0 \\
0 & 0 & 0 \\
\end{bmatrix},
\lambda_2=    \begin{bmatrix}
0 & -i & 0 \\
i & 0 & 0 \\
0 & 0 & 0 \\
\end{bmatrix},
\lambda_3=    \begin{bmatrix}
1 & 0 & 0 \\
0 & -1 & 0 \\
0 & 0 & 0 \\
\end{bmatrix},\nonumber\\
\lambda_4&=    \begin{bmatrix}
0 & 0 & 1 \\
0 & 0 & 0 \\
1 & 0 & 0 \\
\end{bmatrix},
\lambda_5=    \begin{bmatrix}
0 & 0 & -i \\
0 & 0 & 0 \\
i & 0 & 0 \\
\end{bmatrix},
\lambda_6=    \begin{bmatrix}
0 & 0 & 0 \\
0 & 0 & 1 \\
0 & 1 & 0 \\
\end{bmatrix},\nonumber\\
\lambda_7&=    \begin{bmatrix}
0 & 0 & 0 \\
0 & 0 & -i \\
0 & i & 0 \\
\end{bmatrix},
\lambda_8=    \frac{1}{\sqrt{3}}\begin{bmatrix}
1 & 0 & 0 \\
0 & 1 & 0 \\
0 & 0 & -2 \\
\end{bmatrix}.
\end{align}
Gell-Mann matrices are orthogonal with respect to the Hilbert-Schmidt inner product:
\begin{align}
    \mathrm{Tr}[\lambda_m\lambda_n]&=2\delta_{mn}.
\end{align}
The generators $E_l$ of $SU(N)$ are defined so that they are normalized with respect to the trace norm:
\begin{align}
    E_l&\equiv \frac{1}{\sqrt{2}}\lambda_l\nonumber.
\end{align}

\subsection{Generators of $SU(N)$}\label{appendix:GGMM}
The generators of $SU(N)$ correspond to generalized Gell-Mann matrices.
The generalized Gell-Mann matrices are composed of off-diagonal matrices $f_{k,l}$ and diagonal matrices $h_{k,l}$.
\begin{align}
    \lambda_{j,k} &\equiv
\begin{cases}
    O_{k,j}+O_{j,k} & \text{for $1\leq j<k\leq N$},\\
     -i(O_{j,k}-O_{k,j}) & \text{for $1\leq k<j \leq N$},\\
     \sqrt{\frac{2}{j(j+1)}}\left(\sum_{l=1}^j O_{l,l}-jO_{j+1,j+1}\right) & \text{for $1\leq j=k \leq N-1$},
     \end{cases}
\end{align}
where $O_{j,k}$ is the matrix with $1$ in the $jk$-th entry and $0$ elsewhere. The generators $E_{j,k}$ are defined so that they are normalized with respect to the trace norm:
\begin{align}
    E_{j,k}&\equiv \frac{1}{\sqrt{\mathrm{Tr}[\lambda_{j,k}^2]}}\lambda_{j,k}\nonumber,\\
\end{align}

\section{Details of Lie algebraic phase reduction}
To enhance the reader's comprehension, we will elucidate the derivation and technical details of the proposed approach. 
Regarding the derivation of the proposed method, we provide a thorough explanation of the validity of the definition of a limit cycle~[Eq.\limcyc], the reason why the SSE and perturbations can be described using unitary transformations.
As for the technical details, we explain the application of the proposed method in high-dimensional systems, the detailed calculation of perturbations in the phase equations, the correspondence of the limit cycles of the semiclassical and proposed methods in the classical limit. Furthermore, we will elaborate on the importance of considering quantum trajectories in a continuous measurement scheme.

\subsection{Validation for definition of limit cycle in SSE}\label{appendix:Str_valid}
As mentioned in the main text, we require the limit-cycle solution of the SSE to be a pure-state trajectory that is physically observable, in the same manner as the trajectories represented by the SSE.
Then, the limit-cycle dynamics needs to satisfy the norm preservation $\|\ket{\psi}\|=1$.
Since generally a limit-cycle dynamics is represented by ordinary differential equations, we have to extract the deterministic terms from the SSE as its limit-cycle dynamics.
However, the deterministic terms of stochastic differential equations, which include the SSE, vary depending on the stochastic calculus used to describe them.
Hence, we need to define limit-cycle dynamics in the stochastic calculus where the norm is preserved in the deterministic term of the SSE. 

Considering an arbitrary stochastic calculus, stochastic terms of the SSE are evaluated at an arbitrary time $t+pdt$ during $[t,t+dt]$, where $0\leq p\leq 1$ ($p=0$ in the Ito calculus and $p=1/2$ in the Stratonovich calculus).
When the stochastic terms are evaluated at time $t+pdt$, the SSE is represented as follows (terms less than $O(dt)$ are neglected in the following):
\begin{align}
    d\ket{\psi(t)}&=\left[-iH_{\mathrm{eff}}+\sum_{k=1}^M \frac{1}{2}\braket{X_k}L_k - \frac{1}{8}\braket{X_k}^2\right]\ket{\psi(t+pdt)}dt\nonumber\\
    &-p\left[L_k^2-\braket{X_k}L_k+\frac{3}{4}\braket{X_k}^2-\frac{1}{2}\braket{X_kL_k}-\frac{1}{2}\braket{L_k^{\d}X_k}\right]\ket{\psi(t+pdt)}dt\nonumber\\
    &+(L_k-\frac{\braket{X_k}}{2})\ket{\psi(t+pdt)}dW_k(t).
    \label{eq:SSE_p}
\end{align}
The time derivative of the norm $|\braket{\psi|\psi}|$ under the deterministic terms of Eq.~\eqref{eq:SSE_p} can be expressed as follows:
\begin{align}
    d(\braket{\psi|\psi})&=(\bra{\psi}+d\bra{\psi})(\ket{\psi}+d\ket{\psi})-\braket{\psi|\psi}\nonumber\\
    &=(d\bra{\psi})\ket{\psi}+\bra{\psi}(d\ket{\psi})+(d\bra{\psi})(d\ket{\psi})\nonumber\\
    &=\sum_{k=1}^M\left[\frac{1}{4}\braket{X_k}^2-\braket{L_k^{\d}L_k}-p\left(\frac{1}{2}\braket{X_k}^2-2\braket{L_k^{\d}L_k}\right)\right]dt,
\end{align}
Clearly, the norm preservation $d|\braket{\psi|\psi}|=0$ for an arbitrary state $\ket{\psi}$ is satisfied if and only if $p=1/2$, i.e., in the Stratonovich calculus.
Therefore, even when considering an arbitrary stochastic calculus, we should define the limit-cycle dynamics based on the Stratonovich calculus.

\subsection{Hermitian representation of SSE}\label{appendix:Hermitian}
Although the SSE is not represented as an Hermitian dynamics, it can also be represented as an Hermitian dynamics.
The seeming contradiction in this statement arises from the fact that the SSE is a nonlinear equation, unlike the Schrödinger equation, which is linear.
In general, a pure-state dynamics can be represented by a linear operator as follows:
\begin{align}
    d\ket{\psi}=-iA\ket{\psi}dt,
    \label{eq:-iA}
\end{align}
where $A$ is an arbitrary linear operator. 
Under Eq.~\eqref{eq:-iA}, the derivative of the norm $\braket{\psi|\psi}$ can be expressed as follows:
\begin{align}
    d(\braket{\psi|\psi})&=(\bra{\psi}+d\bra{\psi})(\ket{\psi}+d\ket{\psi})-\braket{\psi|\psi}\nonumber\\
    &=d(\bra{\psi})\ket{\psi}+\bra{\psi}d(\ket{\psi})+d(\bra{\psi})d(\ket{\psi})\nonumber\\
    &=-i\braket{(A-A^{\d})}dt,    
\end{align}
where terms smaller than $O(dt)$ are neglected.
If $A$ does not depend on the state $\ket{\psi}$, the norm preservation of pure-state dynamics for the arbitrary state $\ket{\psi}$ is satisfied if and only if $A=A^{\d}$, i.e., Hermitian dynamics.
However, if $A$ does depend on the state $\ket{\psi}$, that is, when Eq.~\eqref{eq:-iA} is a nonlinear equation, it is possible for the norm preservation condition to be satisfied even if $A$ is not an Hermitian operator as Eq.~\limcyc{}.
In fact, the norm-preserved non-Hermitian dynamics including Eq.~\limcyc{} can be represented also by an Hermitian operator as follows:
\begin{align}
\label{eq:A-braketA}
    -i\left[A(\ket{\psi})-\braket{A(\ket{\psi})}\right]\ket{\psi}dt&=-iH_A(\ket{\psi})\ket{\psi}dt,\\
    H_A(\ket{\psi})&\equiv \left[A(\ket{\psi})-\braket{A(\ket{\psi})}\right]\ket{\psi}\bra{\psi}+\mathrm{H.c.},
\end{align}
where $H_A(\ket{\psi})$ is an Hermitian operator and the difference between Eqs.~\eqref{eq:-iA} and \eqref{eq:A-braketA}, i.e., the term $-i\braket{A}\ket{\psi}dt$, can be neglected since it corresponds to a $U(1)$ transformation, which has no physical effect on the state $\ket{\psi}$.
The same statement holds true for the Stratonovich calculus.
Although the SSE and Eq.~\limcyc{} are not described as Hermitian dynamics, they satisfy norm preservation and represent pure-state dynamics due to their non-linearity. Thus, they can also be represented as unitary transformations by the Hermitian operators $H_A(\ket{\psi})$.
Furthermore, decomposing the Hermitian operator $H_A(\ket{\psi})$ into generators of a Lie algebra on the limit-cycle solution $\ket{\psi_0(\theta)}$, we can calculate the phase response curve (PRC) for arbitrary weak perturbations as long as its dynamics is described by the SSE.

\subsection{Conversion of stochastic terms of SSE into Hermitian operators}\label{appendix:derivation}
In the main paper, the PRC is calculated with respect to unitary transformation perturbations.
Although Eq.~\eqref{eq:SSE_Str} is not formulated by the Hermitian operators, an arbitrary infinitesimal change of a pure state can be represented as a unitary transformation. Therefore, Eq.~\eqref{eq:SSE_Str} and its deterministic term should be represented as unitary transformations because of their norm-preserving properties. Since, by the phase definition the deterministic terms in Eq.~\eqref{eq:SSE_Str} correspond to a constant $\omega$ in the phase space, we focus on the stochastic terms. The stochastic terms can be decomposed into the components parallel and orthogonal to $\ket{\psi}$:
\begin{align}
    d\ket{\psi}&=\sum_{k=1}^M \left(L_k -\frac{\braket{X_k}}{2}\right)\ket{\psi}\circ dW_k \nonumber\\
    &=\sum_{k=1}^{M}\left(\frac{1}{2}\braket{L_k-L_k^{\d}}\ket{\psi}+(L_k-\braket{L_k})\ket{\psi}\right)\circ dW_k\nonumber\\
    &=\sum_{k=1}^{M}\left(\frac{1}{2}\braket{L_k-L_k^{\d}}\ket{\psi}-iH_k\ket{\psi}\right)\circ dW_k,
    \label{eq:SSE_stoch}
\end{align}
where Hermitian operators $H_k$ are defined as follows:
\begin{align}
    H_k \equiv i(L_k-\braket{L_k})\ket{\psi}\bra{\psi}+\mathrm{H.c.}
    \label{eq:Hk}
\end{align}
In Eq.~\eqref{eq:SSE_stoch}, from the first to the second line, we decompose the stochastic terms into the parallel and orthogonal components to state $\ket{\psi}$. The parallel components correspond to $U(1)$ transformation and have no effect on the phase. From the second to the third line, we represent the orthogonal components by traceless Hermitian operators $H_k$, which are defined in Eq.~\eqref{eq:Hk}. Traceless Hermitian operators $H_k$ perform rotation in the plane spanned by $\ket{\psi}$ and $(L_k-\braket{L_k})\ket{\psi}$. By the trace-orthogonal property of the Lie algebra, traceless Hermitian operators $H_k$ can be decomposed into a linear combination of $SU(N)$ generators as follows:
\begin{align}
    H_k&=\sum_{l=1}^{N^2-1}g_{k,l}E_l,\label{eq:Hk_def}\\
    g_{k,l}&\equiv \mathrm{Tr}[H_{k} E_l].\label{eq:gkl_def}
\end{align}
Similarly, the deterministic terms in Eq.~\eqref{eq:SSE_Str} are represented by a traceless Hermitian operator and decomposed into generators of a Lie algebra with coefficients.

\subsection{Calculation of PRC for high-dimensional systems}\label{appendix:high_dim}
For high-dimensional systems, the calculation of PRCs with respect to $N^2-1$ Lie algebra generators is computationally demanding. In such a case, we can directly calculate the PRC $Z_p(\theta)$ for arbitrary Hermitian operator $H_p$ using a direct method with the isochron $\Theta(\ket{\psi})$ as follows:
\begin{align}
    Z_p(\theta)\equiv\lim_{\epsilon\to 0}\frac{\Theta(\exp(-i\epsilon H_p)\ket{\psi_0(\theta)})-\theta}{\epsilon}.
\end{align}
Alternatively, considering the norm-preservation $|\ket{\psi}|=1$ and $U(1)$ symmetry, we can project N-dimensional complex vector $\ket{\psi}$ to $2N-2$ real vector $v$ as follows:
\begin{align}
\ket{\psi}&=\begin{pmatrix}
    r_1e^{i\theta_1}\\r_2 e^{i\theta_2}\\\vdots\\r_N e^{i\theta_N},
\end{pmatrix},\\
\theta'_i&\equiv \theta_{i+1}-\theta_N,\\
v&\equiv(r_1,\cdots, r_{N-1},\theta'_1,\cdots,\theta'_{N-1})^T,
\end{align}
where $r\geq 0$ and $0\leq\theta'<2\pi$. Employing the direct or adjoint method~\cite{10.1162/neco.1996.8.5.979}, we can obtain PRCs with respect to the basis vectors of $v$. While the formulation with respect to $2N-2$ basis vectors is of lower dimension than that with respect to $N^2-1$ Lie algebra generators, it requires the calculation for $d\ket{\psi}=-i\epsilon H_p(t)\ket{\psi_0(\theta_0(t))}$ for arbitrary unitary perturbation $\exp(-i\epsilon H_p(t))$, where $0<\epsilon \ll 1$. Therefore, this formulation is not completed in the phase space without limit-cycle solution $\ket{\psi_0(\theta)}$, unlike that with respect to a Lie algebra.

\subsection{Perturbation in phase equation}\label{appendix:details}
When a weak perturbation Hamiltonian is added, the quantum limit cycle model is described by the Lindblad equation as follows:
\begin{align}
    \frac{d\rho}{dt}=-i[H+\epsilon H_p,\rho]+\sum_{k=1}^M\mathcal{D}[L_k]\rho,
    \label{eq:perturb}
\end{align}
where $H_p$ is the perturbation Hamiltonian and $\epsilon$ is the perturbation strength. Here, we assume that the perturbation is sufficiently weak; that is, $\epsilon\ll1$.
In the absence of perturbation, that is, $H_p=0$, the limit cycle solution $\ket{\psi_0}$ is defined under the deterministic terms of the SSE [Eq.~\limcyc] derived from Eq.~\eqref{eq:perturb}.
Therefore, in the case of strong squeezing, the limit cycle is shown in Fig.~\FIGWigner(b), where squeezing is included in the system Hamiltonian $H$, not in $H_p$.
According to Eq.~\PEUstr, Eq.~\eqref{eq:perturb} is converted into the following phase equation:
\begin{align}
    \frac{d\theta}{dt}=\omega+ \epsilon\sum_{l=1}^{N^2-1}f_{p,l}Z_l(\theta) +\sum_{k=1}^M \sum_{l=1}^{N^2-1}g_{k,l}(\theta)Z_l(\theta) \circ \xi_k(t),
    \label{eq:perturbPE}
\end{align}
where $f_{p,l}$ denotes a coefficient defined as $f_{p,l}\equiv\mathrm{Tr}[H_p E_l]$.
The phase distribution shown in Fig.~\FIGdistUprc~is obtained from the perturbed phase equation [Eq.~\eqref{eq:perturbPE}], with $H_p=-i(a-a^{\dagger})$.
Furthermore, the PRC $Z_p(\theta)$ with respect to $H_p=-i(a-a^{\dagger})$ in Fig.~\FIGdistUprc~was obtained from a linear combination of the PRCs [Eq.~\phaseresponse] as $Z_p(\theta)=\sum_{l=1}^{N^2-1}f_{p,l}Z_l(\theta)$ or the direct method as $Z_p(\theta)=\mathrm{lim}_{h\to 0}[\Theta(\mathrm{exp}(-ih H_p \ket{\psi_0(\theta)})-\Theta(\ket{\psi_0(\theta)}))]/h$.
As discussed in Sec.~\ref{appendix:derivation} of the Supplementary Material, based on the homodyne detection scheme, the perturbation can be considered not only in the Hamiltonian form $\epsilon H_p$, but also dissipator form $\epsilon \mathcal{D}[L_p]\rho$.  

\subsection{Correspondence of limit cycles of semiclassical and proposed methods in classical limit}
In the semiclassical method, though the Lindblad equation needs to be converted into the Fokker-Planck equation of quasiprobability distribution P-representation, this conversion is not always available.
According to the standard techniques~\cite{gardiner2004quantum}, the relationship between the P-representation $P(\alpha,\alpha^*)$ and the density operator $\rho$ is described as follows:
\begin{align}
    a\rho &\to \alpha P(\alpha,\alpha^*), \quad a^{\dagger}\rho \to (\alpha^*-\frac{\partial}{\partial \alpha}) P(\alpha,\alpha^*),\quad 
    \rho a \to (\alpha-\frac{\partial}{\partial \alpha^*}) P(\alpha,\alpha^*), \quad \rho a^{\dagger} \to \alpha^* P(\alpha,\alpha^*),
    \label{eq:standard}
\end{align}
where $a$ and $a^{\dagger}$ are creation and annihilation operators, respectively. 
According to Eq.~\eqref{eq:standard} and the definition of the Lindblad dissipator $\mathcal{D}[L]\rho\equiv L\rho L^{\dagger}-(1/2)(L^{\dagger}L\rho+\rho L^{\dagger}L)$, the dissipators of $a^3$, $a^{\dagger 2}$, and the higher terms correspond to the third or higher derivatives of P-representation.
Because the Fokker-Planck equation is composed of the second and lower derivatives of the P-representation, these terms are not applicable in the semiclassical method without approximation techniques.
Therefore, we consider the case where jump operators are defined as linear combination of $\{I,a,a^2\}$ or $\{I,a,a^{\dagger},a^{\dagger}a\},$.

In this case, the drift terms of the semiclassical Fokker-Planck equation are equivalent to the deterministic terms of the SSE in the classical limit.
In other words, the Lindblad equation composed of these dissipators has the same limit cycle for both the semiclassical and proposed methods in the classical limit. 

\subsection{Importance of continuous measurement scheme}
In quantum dynamics, the density operator cannot be observed directly, and determining the density operator requires many measurement outcomes. However, performing measurements on a quantum system can interfere with and alter the density operator. Therefore, a non-demolition measurement, in which the density operator remains unchanged before and after the measurement, is necessary to determine its time evolution. Non-demolition measurements can be implemented both in direct and indirect measurements. Direct non-demolition measurements can be achieved when the density operators are measured by the positive operator-valued measure $\Pi\equiv\sum_k\ket{\lambda_k}\bra{\lambda_k}$, where $\ket{\lambda_k}$ represents eigenvector of the density operator as $\rho=\sum_k p_k\ket{\lambda_k}\bra{\lambda_k}$. This approach is somewhat self-contradictory and not entirely effective because knowledge of the eigenvectors of the density operators is required to determine the operators. On the other hand, indirect non-demolition measurements can be achieved through the continuous measurement schemes that are widely implemented in experimental settings.
Although the semi-classical Langevin equations and quantum Monte Carlo methods can generate trajectories equivalent to the time evolution of density operators, they do not consider measurements. Consequently, the physical interpretation and implementation of these trajectories is challenging. Therefore, continuous measurement serves as the necessary methodology for obtaining the time evolution of the density operator.

\section{Application to finite-level systems}
We will present several examples as a demonstration of the applicability of our proposed method to finite-level systems.
We detail the systems of qubits, spin-$1$, and spin-$3/2$ that are used in the main text, along with the parameter settings used in Fig.~3. 
In addition, we also introduce two quantum oscillators, a qubit under bit-flip error and a $\Lambda$ atom.

\subsection{Qubit system}\label{appendix:qubit_limcyc}
Let us consider the Lindblad equation of a two-level system~\cite{PhysRevA.101.062104} as follows:
\begin{align}
    \frac{d\rho}{dt}=-i[\Delta S_z,\rho]+\gamma_+\mathcal{D}[S_+]\rho+\gamma_-\mathcal{D}[S_-]\rho,
    \label{eq:tls}
\end{align}
where the spin operators $S_{x,y,z}$ are the generators of the rotation group $SO(2)$, the pumping and damping operators $S_{\pm}$ are defined by $S_{\pm}=(1/\sqrt{2})(S_x\pm iS_y)$, $\Delta$ is the detuning of the system, $\gamma_{\pm}$ are the pumping and damping rates.
In the homodyne detection scheme, we can derive the diffusive SSE [Eq.~2] from Eq.~\eqref{eq:tls}, and it can form the limit cycle.
In Fig.~3(a), the parameters are set as $(\Delta$, $\gamma_+$, $\gamma_-)=(3.0$, $0.1$, $0.05)$.

\subsection{Spin-$1$ system}\label{appendix:spin1_limcyc}
Let us consider the Lindblad equation of a three-level spin-1 system as follows:
\begin{align}
    \frac{d\rho}{dt}=-i[\Delta S_z,\rho]+\gamma_+(1+n_+)\mathcal{D}[S_+S_z]\rho+\gamma_-(1+n_-)\mathcal{D}[S_-S_z]\rho+\gamma_+n_+\mathcal{D}[S_zS_-]\rho+\gamma_-n_-\mathcal{D}[S_zS_+]\rho,
    \label{eq:spin1}
\end{align}
where the spin operators $S_{x,y,z}$ are the generators of the rotation group $SO(3)$, the pumping and damping operators $S_{\pm}$ are defined by $S_{\pm}=(1/\sqrt{2})(S_x\pm iS_y)$, $\Delta$ is the detuning of the system, $\gamma_{\pm}$ are the pumping and damping rates, and the Bose-Einstein distribution $n_{\pm}$ is defined as $n_{\pm}=(\exp(\omega_{\pm}/T_{\pm})-1)^{-1}$ with the temperature $T_\pm$ and the pumping and damping frequencies $\omega_{\pm}$.
In the zero-temperature limit $T_{\pm}\to 0$, Eq.~\eqref{eq:spin1} corresponds to the model in~\cite{PhysRevLett.121.053601}, where $n_\pm$ and the corresponding inverse processes vanish.
However, in the zero-temperature limit, Eq.~\eqref{eq:spin1} converges on the steady pure state $\ket{1}$ and does not oscillate. At a finite temperature $T_{\pm}>0$, the SSE of Eq.~\eqref{eq:spin1} can form the limit cycle. 
In Fig.~3(b), the parameters are set as $(\Delta$, $\gamma_+$, $\gamma_-$, $n_{+}$, $n_{-})=(2.0$, $0.01$, $0.005$, $0.2$, $0.3)$.

\subsection{Spin-$3/2$ system}\label{appendix:spin32_limcyc}
Let us consider the Lindblad equation of a spin-$3/2$ system as follows:
\begin{align}
    \frac{d\rho}{dt}=&-i\left[\Delta S_z,\rho\right]+\gamma_+\mathcal{D}[S_+S_z]\rho+\gamma_-\mathcal{D}[S_-S_z]\rho,
    \label{eq:spin32}
\end{align}
where the spin operators $S_{x,y,z}$ are the generators of the rotation group $SO(4)$, the pumping and damping operators $S_{\pm}$ are defined by $S_{\pm}=(1/\sqrt{2})(S_x\pm iS_y)$, $\Delta$ is the detuning of the system, $\gamma_{\pm}$ are the pumping and damping rates.
Unlike the spin-1 system, the half-integer system does not converge to a pure steady state under the Lindblad equation [Eq.~\eqref{eq:spin32}]; hence, the proposed method can be applied without considering the reverse process.
In Figs.~3(c) and (d), the parameters are set as $(\Delta$, $\gamma_+$, $\gamma_-)=(2\pi$, $1.0$, $0.1)$.

\subsection{Qubit system under bit-flip error}\label{appendix:qubit_limcyc2}
Let us consider the Lindblad equation of a qubit system with bit-flip noise as follows:
\begin{align}
    \frac{d\rho}{dt}=-i[\Omega \sigma_z,\rho]+\gamma\mathcal{D}[\sigma_x]\rho,
    \label{eq:refB:Lindbit}
\end{align}
where $\sigma_s$ is the Pauli-s matrix, $\Omega$ is the natural frequency of the system, and $\gamma$ is the bit-flip error rate. In the homodyne detection scheme, the diffusive SSE corresponding to Eq.~\eqref{eq:refB:Lindbit} is as follows:
\begin{align}
    d\ket{\psi}=\left[-i\omega\sigma_z+\gamma\braket{\sigma_x}\sigma_x-\gamma\braket{\sigma_x}^2\right]\ket{\psi}dt+\sqrt{\gamma}(\sigma_x-\braket{\sigma_x})\ket{\psi}\circ dW,
\label{eq:refB:SSEbit}
\end{align}
where the related homodyne current is $J\equiv\sqrt{\gamma}\braket{\sigma_x}+\xi(t)$.
We can provide a simple explanation for this. The state of a qubit corresponds to a point on or inside the Bloch sphere. 
As the system approaches a steady state, the state of the qubit moves to the XY-plane.
As a result, the SSE [Eq.~\eqref{eq:refB:SSEbit}] yields a limit-cycle solution on the XY-plane.

\subsection{$\Lambda$ atom}\label{appendix:Lambda_limcyc}
Let us consider the Lindblad equation of a three-level $\Lambda$ atom~\cite{PhysRevA.91.062308} as follows:
\begin{align}
    \frac{d\rho}{dt}=-i\left[\sum_{i=0}^2\omega_i\ket{i}\bra{i},\rho\right]+\gamma_1\mathcal{D}[\cos{\phi}\ket{0}\bra{2}+e^{i\eta}\sin{\phi}\ket{1}\bra{2}]\rho+\gamma_2\mathcal{D}[\cos{\alpha}\ket{0}\bra{1}+\sin{\alpha}\ket{1}\bra{0}]\rho,
\end{align}
where $\omega_i$ is the characteristic frequency, $\gamma_1$ is the jump rate from $\ket{2}$ to $\ket{0}$ and $\ket{1}$, $\gamma_2$ is the jump rate between $\ket{0}$ and $\ket{1}$, and $\phi$, $\eta$, and $\alpha$ are the phase parameters of each jump event.
Under continuous measurement, this quantum oscillator oscillates between the two lower levels as the highest level becomes empty.
%